\documentclass[a4paper,11pt]{article}
\usepackage{pos}

\title{Towards the Application of Skewed
Detailed Balance in Lattice Gauge
Theories}
\author{Marina Krstic Marinkovic}
\author*{Joao C. Pinto Barros}

\affiliation{Institute for Theoretical Physics, ETH Zürich,\\
Wolfgang-Pauli-Str. 27, 8093 Zürich, Switzerland}

\emailAdd{marinama@phys.ethz.ch}
\emailAdd{jpinto@phys.ethz.ch}

\abstract{State-of-the-art algorithms in lattice gauge theory typically rely heavily on detailed balance, which is an instrumental tool to prove the correct convergence of the Markov Chain Monte Carlo Algorithm. In this work, we investigate an alternative to detailed balance, skewed detailed balance, and the prospects of alleviating the topological freezing problem by studying the one-dimensional $O\left(2\right)$ model.}

\FullConference{
  The 38th International Symposium on Lattice Field Theory,\\
  August 8 to August 13, 2022\\
  Bonn, Germany
}

\begin{document}
\maketitle

\section{Introduction}

In a Markov Chain Monte Carlo (MCMC) algorithm, ergodicity and detailed balance guarantee the convergence to the desired probability distribution. While ergodicity ensures its existence, detailed balance guarantees that the distribution is the target one. Thus, convergence does not necessitate detailed balance. MCMC simulations of Quantum Chromodynamics (QCD), and other theories, suffer from topological freezing (see e.g. \cite{alles1996,del2004,schaefer2011,mcglynn2014,bonati2018}), which is an instance of critical slowing down. This constitutes an effective ergodicity breaking. While, typically, the algorithm remains ergodic, the time necessary to sample all relevant configurations grows increasingly large. Ergodicity is then effectively broken and, with it, the guarantee that the obtained results can be trusted.

%The problem is tied to the existence of different topological sectors and the difficulty of tunneling between them. In the continuum, each configuration can be associated to a different integer corresponding to a topological charge $Q\in\mathbb{Z}$. Each sector has instanton solutions of the classical equations of motion, which are minima of the action. Instantons characterized by different topological charge cannot be continuously deformed into each other, which results in an infinite action barrier between them. As a result, it is not possible to change the topological sector by local changes of the configuration, at least when approaching the continuum limit.

%Detailed balance and ergodicity are typical requirements of a standard MCMC algorithm. It may be then surprising, at first glance, that one could get in the way of the other. Detailed balance is a strong condition on the structure of the transition matrix of an algorithm, but it is by no means necessary for the algorithm to be correct. 
Applying detailed balance breaking algorithms has been proven advantageous in various situations\footnote{Perhaps the most well-known example of detailed balance breaking is sequential updating in classical spin systems. When applying the Metropolis algorithm, for example, one can construct an algorithm that chooses a random spin to be updated at every step or, alternatively, an algorithm that transverses the lattice sequentially. The latter breaks detailed balance but, when done correctly, still converges to the correct distribution and is more efficient.} (\cite{diaconis2000,bernard2009,suwa2010,schram2015,kaiser2017,faizi2020} to name a few).
It can be argued intuitively that it is expected that algorithms breaking detailed balance can outperform those that do not. Detailed balance corresponds to an equilibrium condition where all elementary processes are in equilibrium with their inverse.  In MCMC algorithms this implies that it is possible that for large Monte Carlo times, the algorithm keeps coming to the same configuration, or remains in a certain "neighborhood", not exploring fast enough the full configuration space. In turn, if one allows for detailed balance-breaking processes, one may forbid the updates to flow in a preferred direction in the configuration space, exploring it faster. This idea is further explored below.

While the ultimate goal is to address this problem in QCD, $O\left(N+1\right)$ models in $N$ dimensions are an interesting setting to explore. While being much simpler, they also have non-trivial topological sectors. On the other side, MCMC algorithms that sample effectively the different topological sectors are well known, as the Wolff cluster algorithm \cite{wolff1989,wolff1990}. Here we will be focusing on a type of topological action \cite{bietenholz2010} for the 1-d $O\left(2\right)$ model.

\section{A 1-d \texorpdfstring{$O\left(2\right)$} Topological Action \label{1do2}}

The 1-d $O\left(2\right)$ model is a simple model exhibiting different topological sectors. Throughout the text, we shall always assume periodic boundary conditions. Under those circumstances and in the continuum, the action can be regarded as a function that takes functions that map the circle on itself: $\theta:O\left(2\right)\rightarrow O\left(2\right)$. Finite action configurations satisfy $\theta\left(x+2\pi\right)=\theta\left(x\right)+2\pi Q$, for some integer $Q$, where $Q$ represents the topological charge and corresponds to the winding number of the $O\left(2\right)$ vector for a given configuration. As it is not possible to continuously deform $\theta\left(x\right)$ with a given winding $Q$ to a new configuration $\theta^\prime\left(x\right)$ with a different winding $Q^\prime\neq Q$, these configurations can be said to be separated by infinite action barriers. 

On the lattice, the standard action takes the form $S_\mathrm{std}=-1/g^2\sum_{n=1}^N\cos\left(\theta_{n+1}-\theta_n\right)$. While the application of local algorithms, like Metropolis-Hastings or heat-bath, works in a very similar way as long as the action remains local, some technical difficulties arise when applying skewed detailed balance to lattice field theories of continuous variables. As a testing ground, we use a topological lattice action \cite{bietenholz2010}
\begin{equation}
    S=\sum_{n=1}^N s_\delta\left(\theta_{n+1}-\theta_n\right), 
    \quad
    \quad
    %\mathrm{where\ }
    s_\delta\left(\Delta\right)=\left\{
    \begin{array}{ll}
          0\  & \cos\left(\Delta\right)<\delta \\
          s_0\  & \mathrm{otherwise}
    \end{array} 
    \right. .
    \label{topo_s_1}
\end{equation}
It strips down the 1-d $O\left(2\right)$ model to the basic features we want to explore. The parameter $\delta$ plays the role of the coupling ($g$ in the standard action). In turn, $s_0$ is the height of the barriers that separate different topological sectors. For a system of size $L$, configurations of the form $\theta=\theta_0+2\pi nQ/L$ have topological charge $Q$ and action $0$, as long as the system is large enough so that $\delta<2\pi Q/L$. When $\delta$ is sufficiently small, it is impossible to continuously\footnote{Here "continuous" means "changing one spin at a time".} deform the action and change topological charge without having an action equal to $s_0$, at least once. In \cite{bietenholz2010} this parameter was set to $s_0\rightarrow\infty$, creating an infinite barrier between different topological sectors. While topological freezing usually arises because of an \emph{effective} breaking of ergodicity, local algorithms like Metropolis-Hastings and heat-bath break \emph{explicitly} ergodicity, when the barrier is infinite.  The present formulation of SDB is designed to accelerate the mixing of local update algorithms and will not be able to deal with that problem. As a consequence, we choose $s_0$ to be finite making barriers between topological sectors potentially large, but not infinite. 

\section{Skewed Detailed Balance - SDB}

\subsection{Global Balance and Detailed Balance}

Let us denote the probability of a given configuration $c$ by ${\cal P}\left(c\right)$ and the transition probabilities, defining a given MCMC algorithm, for the probability of transitioning from a configuration $c$ to $c^\prime$, to be $T\left(c\rightarrow c^\prime\right)$. In order to simulate correctly the model, we need to guarantee that the stationary distribution of the Markov chain corresponds to our target probability distribution,
\begin{equation}
    {\cal P}\left(c\right)
    =\sum_{c^\prime}
    {\cal P}\left(c^\prime\right)T\left(c^\prime\rightarrow c\right),
    \label{global_balance}
\end{equation}
which we shall refer as \emph{global balance}.
Proving that a given transition matrix $T$ satisfies the above condition is a complicated task, in general. In contrast, \emph{detailed balance} condition constitutes a more stringent requirement

\begin{equation}
    {\cal P}\left(c\right)T\left(c\rightarrow c^\prime\right)
    =
    {\cal P}\left(c^\prime\right)T\left(c^\prime\rightarrow c\right).
    \label{detailed_balance}
\end{equation}
Together with ergodicity, it guarantees that the probability distribution simulated is precisely given by ${\cal P}$, which can be easily seen by summing over $c^\prime$ on both sides of Eq. \eqref{detailed_balance}. A natural question is whether we can use the freedom to break this condition to construct more efficient algorithms.

\subsection{Lifting and Skewed Detailed Balance}

Skewed detailed balance condition provides an alternative path to satisfy Eq. \eqref{global_balance}. The first step consists on applying the concept of lifting \cite{diaconis2000,faizi2020,chen1999}, where we create a new replica of the phase space and introduce a \emph{lifting variable} $\varepsilon=\{+,-\}$ that identifies the two copies. They occur with the same probability $\varepsilon$: ${\cal P}\left(c,\varepsilon\right)={\cal P}\left(c\right)/2$. In the enlarged space, global balance reads ${\cal P}\left(c\right)=\sum_{c^\prime,\varepsilon^\prime}{\cal P}\left(c^\prime\right)T\left(c^\prime,\varepsilon^\prime\rightarrow c,\varepsilon\right)$,
We construct a transition matrix dependent on $\varepsilon$ such that the \emph{skewed balance} condition is satisfied

\begin{equation}
    {\cal P}\left(c\right)T\left(c,\varepsilon\rightarrow c^\prime,\varepsilon\right)
    =
    {\cal P}\left(c^\prime\right)T\left(c^\prime,-\varepsilon\rightarrow c,-\varepsilon\right).
    \label{skewed_balance}
\end{equation}
In order to make the algorithm ergodic in the enlarged space, the new variable $\varepsilon$ must also be updated, i.e., there must be non-zero matrix elements of the form $T\left(c,\varepsilon\rightarrow c,-\varepsilon\right)\neq0$. In contrast to the detailed balance condition, Eq. \eqref{skewed_balance}  does not guarantee the correct probability distribution. By imposing the latter and using Eq. \eqref{skewed_balance} on the global balance condition,  we arrive at the condition

\begin{equation}
    T\left(c,\varepsilon\rightarrow c,-\varepsilon\right)
    -
    T\left(c,-\varepsilon\rightarrow c,\varepsilon\right)
    =\sum_{c^\prime\neq c}
    \left(
    T\left(c,-\varepsilon\rightarrow c^\prime,-\varepsilon\right)
    -
    T\left(c,\varepsilon\rightarrow c^\prime,\varepsilon\right)
    \right),
\end{equation}
where we have implicitly assumed that $c$ and $\varepsilon$ are never updated together.
The equation above admits a large range of solutions. Here we will always adopt the Turintsyn-Chertkov-Vucelja type solution \cite{turitsyn2011}, which is given by

\begin{equation}
    T\left(c,\varepsilon\rightarrow c,-\varepsilon\right)
    =
    \max\bigg\{0,
    \sum_{c\neq c^\prime}\left[T\left(c,-\varepsilon\rightarrow c^\prime,-\varepsilon\right)
    -
    T\left(c,\varepsilon\rightarrow c^\prime,\varepsilon\right)\right]
    \bigg\}.
    \label{TCV}
\end{equation}
Note that $\sum_{c\neq c^\prime}T\left(c,\varepsilon\rightarrow c^\prime\varepsilon\right)$ is the probability of updating the configuration $c$ to a different configuration $c^\prime$ at constant $\varepsilon$. With this choice, as long as this probability is larger than its counterpart at constant $-\varepsilon$, the system will remain in $\varepsilon$. This is a nice property of the algorithm: it detects when the acceptance rate is larger in the other replica, and only then there is a finite probability to move there.

\subsection{From Detailed Balance to Skewed Balance: The Skewness Function}

Metropolis or heat-bath type of updates provide a straightforward way to construct algorithms that satisfy detailed balance. A priory, it is not obvious how to construct updates satisfying Eq. \eqref{skewed_balance}. Following \cite{sakai2016}, we can adapt an algorithm satisfying detailed balance to construct an algorithm that satisfies skewed balance on the extended model, after lifting has been applied. In order to see how, assume that we have an algorithm satisfying detailed balance, characterized by a transition matrix $T_\mathrm{DB}$ and a \emph{skewness function} $\vartheta$ satisfying $\vartheta_\varepsilon\left(c,c^\prime\right)=\vartheta_{-\varepsilon}\left(c^\prime,c\right)$.
We can then verify that
\begin{equation}
    T\left(c,\varepsilon\rightarrow c^\prime,\varepsilon\right)
    =
    \vartheta_\varepsilon\left(c,c^\prime\right)
    T_\mathrm{DB}\left(c\rightarrow c^\prime\right)
\end{equation}
satisfies the skewed balance condition Eq. \eqref{skewed_balance}. The task of constructing the algorithm is now shifted to find a suitable skewness function. Naturally, for $T$ to be a proper transition matrix, it should remain positive and smaller or equal to one. A sufficient condition is to impose $0\leq \vartheta_\varepsilon\left(c,c^\prime\right)\leq 1,\ \forall \varepsilon c,c^\prime$. The function $\vartheta$ works as a post selection factor, which will favor certain updates over others, depending on the value of $\varepsilon$. This can be constructed by taking into account the effect of the update on an observable of choice \cite{sakai2013,hukushima2013,faizi2020}. Suppose we take an observable $O$ that, for each configuration $c$, $O\left(c\right)\in\mathbb{R}$. We can then define

\begin{equation}
    \vartheta_\varepsilon\left(c,c^\prime\right)
    =\frac
    {1+\varepsilon\lambda\mathrm{sign}\left(O\left(c^\prime\right)-O\left(c\right)\right)}
    {1+\lambda},
\end{equation}
where we have used the sign function, which returns the sign of its argument, and $\lambda$ is a free parameter that can be used to explore the degree to which we break detailed balance. If $\lambda=0$, the skewness function is always 1, and detailed balance is satisfied. Conversely, if $\lambda=1$ all updates that change the observable in a direction opposed to $\varepsilon$ will get rejected.

This approach has been used, for Ising and Potts models, with the magnetization defining the skewness function \cite{sakai2013,hukushima2013,faizi2020}. The idea consists of constructing an algorithm that can move from smaller to larger values of magnetization by eliminating some of the randomness of the original one. If $\lambda=1$, then we either only accept increasing the magnetization (when $\varepsilon=1$) or decreasing it (when $\varepsilon=-1$) until the acceptance rate has become small and the replica variable is updated. This has been shown to provide an effective advantage.

\section{Instanton-Skewed Detailed Balance - ISDB}

As previously discussed, topological freezing occurs because different topological sectors become separated by increasingly large barriers. Breaking detailed balance may provide a pathway to overcome these barriers: if over a succession of updates, we always move in a specific direction, we may hope to climb a large barrier and update the topological sector. It is worth pointing out how this scenario differs from the one describing magnetization in spin models. The magnetization can be suitably updated locally, which is not the case for the topological charge (except for possible lattice artifacts). Furthermore, different values of magnetization are not separated by large action barriers. 
The most obvious choice of observable for the skewness function would be the topological charge $Q$ or the topological susceptibility $Q^2/V$, where $V$ is the volume of the system. Because we will be using a heat-bath algorithm, which is a local update algorithm, we need a different choice for the skewness function.

\subsection{Instantons and the Skewness Function}

As previously discussed, for the $1-$d $O\left(2\right)$ model, one can define configurations that minimize the action: $\theta_n^Q=nQ\frac{2\pi}{L}$.
We would like to construct a skewness function that promotes the transition between different topological charges. We propose defining a family of observables that measure how close the configuration is to a given $\theta^Q$. We define

\begin{equation}
    O_Q\left(c\right)
    =\sum_n
    \cos\left(\theta_n^Q-\theta_n\right),
\end{equation}
where the values $\theta_n$ correspond to the angular values, in the configuration $c$. We do not expect the algorithm to be crucially dependent on the specific choice of the observable, as long as it constitutes a reasonable measure of how far the two configurations differ from each other. This specific choice is proportional to the (standard) action resulting from the interaction of the system with an instanton in the background. We will investigate the algorithm where the skewness function is defined as 

\begin{equation}
    \vartheta_\varepsilon\left(c,c^\prime\right)
    =\frac
    {1+\varepsilon\lambda\mathrm{sign}\left(O_1\left(c^\prime\right)-O_1\left(c\right)\right)}
    {1+\lambda}.
    \label{instanton_skewed}
\end{equation}
This will amount to promote the algorithm to approach the instanton configuration given by $\theta_n=\theta_n^{Q=1}$ , if $\varepsilon=1$, and $\theta_n=\pi+\theta_n^{Q=1}$ otherwise. We will call this algorithm \emph{instanton-skewed detailed balance} algorithm. The goal of the remainder of this paper is to take a standard heat-bath algorithm (equivalent to set $\lambda=0$ in \eqref{instanton_skewed}), satisfying detailed balance, and compare it to its instanton-skewed detailed balance adaptation with $\lambda=1$.

\section{Updating with ISDB}

The proposed algorithm biases the moves to copy one of two instantons on the background, both with charge 1. This looks unnatural, and one may question if the MCMC algorithm even converges to the Boltzmann distribution defined by the action \eqref{topo_s_1}. As argued in the previous section, since SDB is satisfied (Eq. \eqref{skewed_balance}) and transitions between replicas are suitably chosen (Eq. \eqref{TCV}), this is guaranteed to be the case. In this section, we discuss some of the consequences of this algorithm and how it can constitute the building block of more efficient algorithms. The results presented will focus on the constraint angle $\delta=0.75$, system sizes $L\in\{64,128\}$ and barrier heights $s_0\in\{10,100,1000\}$.

\subsection{Updating the Topological Charge}

\begin{figure}
    \centering
    \includegraphics[width=.99\textwidth]{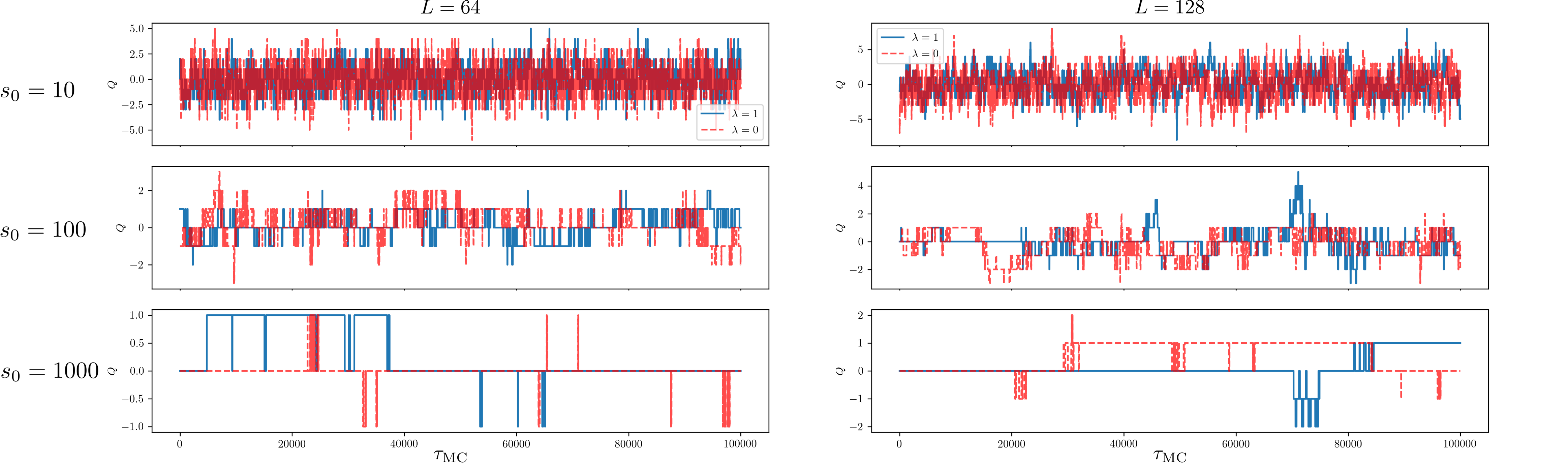}
    \caption{Sampling the topological charge, over Monte Carlo time, for the algorithms that satisfy and break detailed balance. By increasing the action barrier, $s_0$, updating the topological sector becomes less frequent.}
    \label{fig:topo_hist}
\end{figure}

In Section \ref{1do2}, we have argued that $s_0$ represents the height of the barriers that separate topological sectors. The larger its value, the harder will be to update the topological sector. This can be seen explicitly in Fig. \ref{fig:topo_hist}, where we plot the value of the topological charge, after equilibrating, as a function of Monte Carlo time, for different barrier heights $s_0$, and two different volumes.
The plot shows explicitly how increasing the height of the barrier makes it less likely to update the topological charge, both for the standard heat-bath algorithm ($\lambda=0$) and for its detailed balance-breaking version ($\lambda=1$). There is no stark distinction between the algorithms that are exhibited by this plot. One may even expect that the ISDB algorithm will end up performing worse, as it favors transitions between two specific configurations. The main question that we are interested in is whether it can accelerate relaxation when starting near those specific configurations. If so, the principles of construction of this algorithm may be used to construct a more efficient algorithm that does not rely on such specificities.

\subsection{Equilibration of Charged Configurations}

\begin{figure}
    \centering
    \includegraphics[width=.40\textwidth]{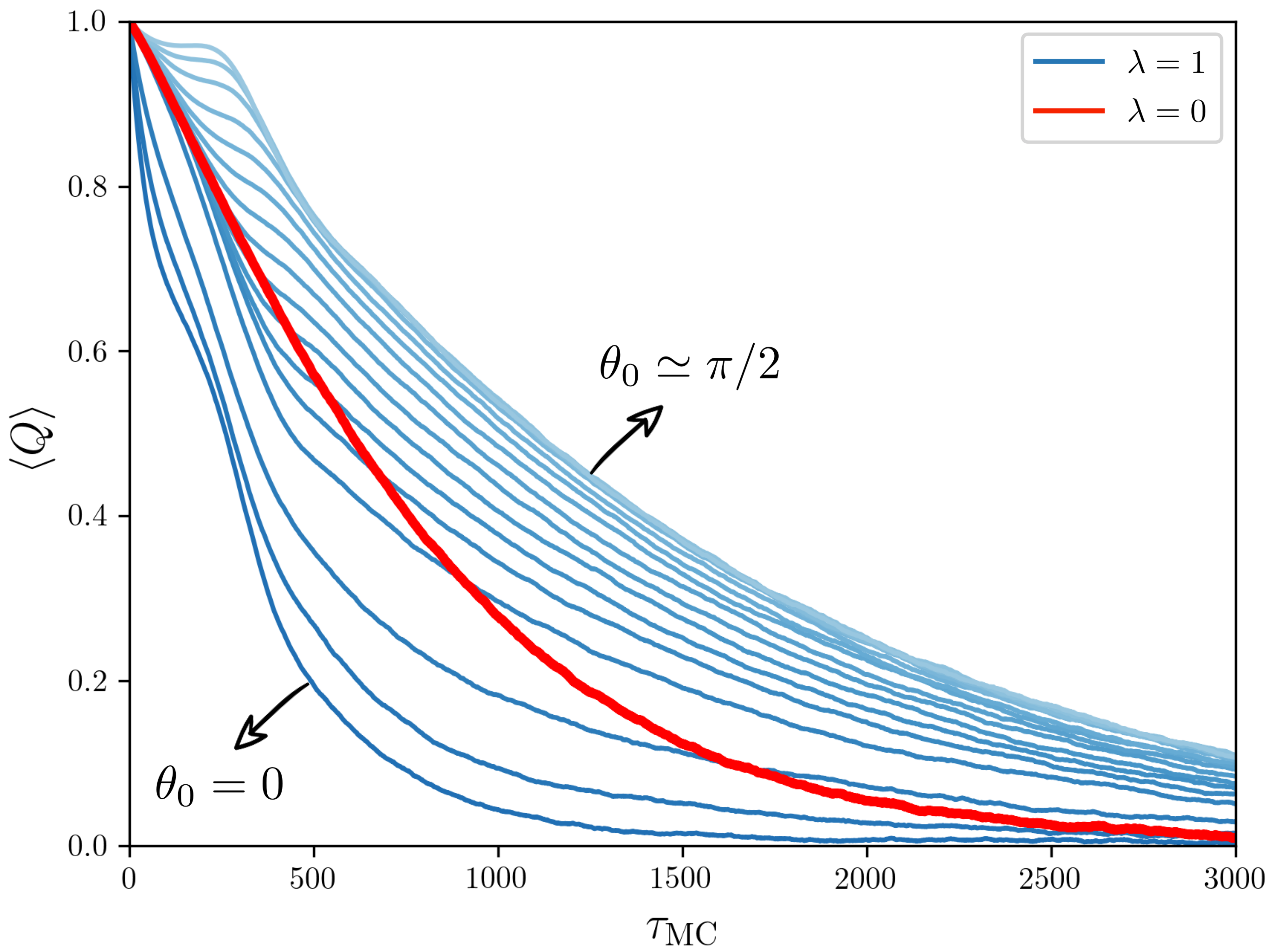}
    \hspace{1cm}
    \includegraphics[width=.40\textwidth]{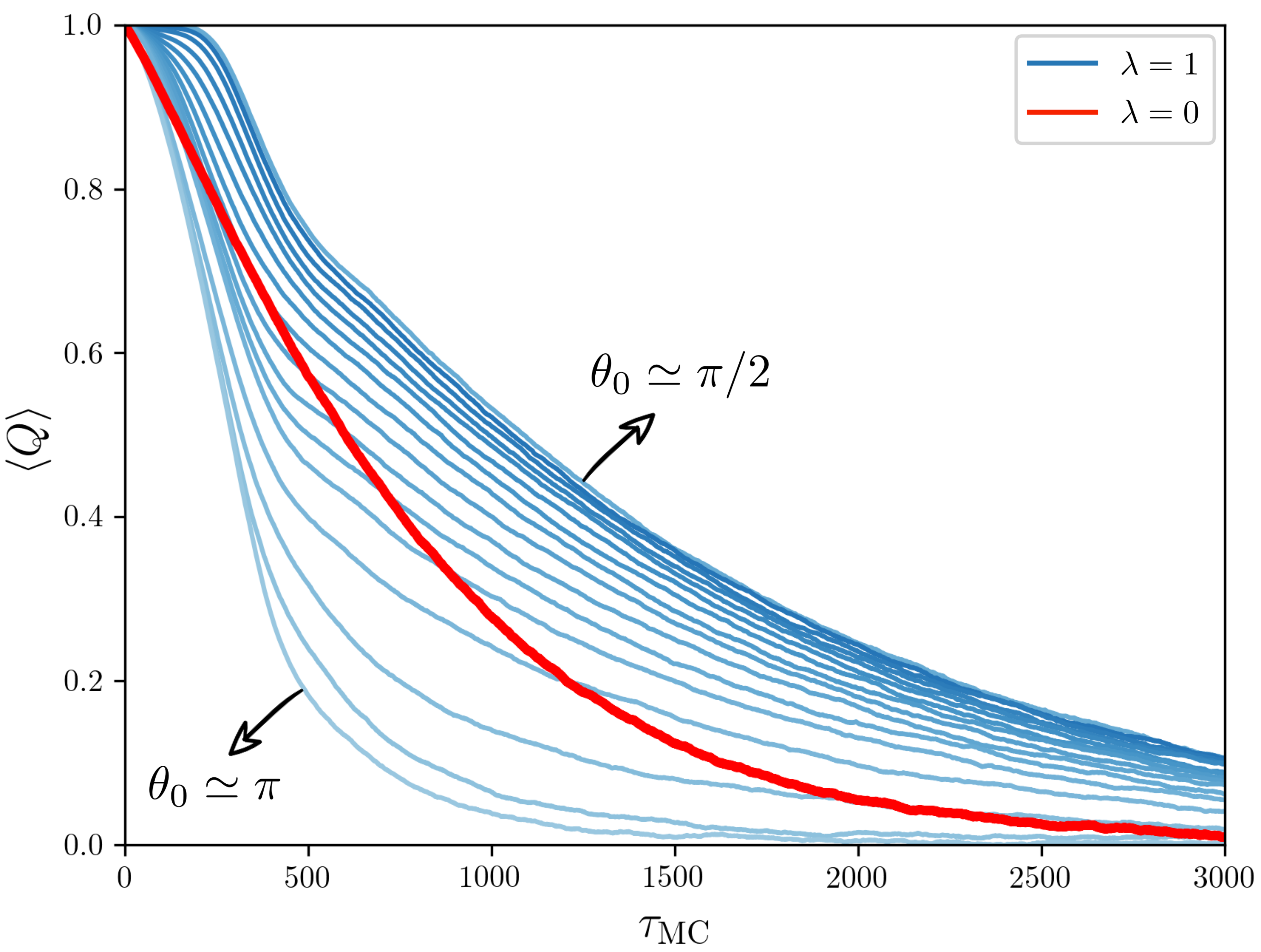}
    \caption{Comparison between the detailed balance breaking algorithm in shades of blue ($\lambda=1$) and its detailed balance counterpart in red ($\lambda=0$) for $L=128$. The different blue lines represent different initial configurations, where $\theta_0$ varies in steps of  $0.1$. The efficiency of the standard algorithm is independent of the initial condition and is represented by a single line. On the left $\theta_0\in\left[0,\pi/2\right]$ and on the right $\theta_0\in\left[\pi/2,\pi\right]$.}
    \label{fig:relaxation}
\end{figure}

In this preliminary stage, we focus in a simple question: how does the system relax, when topological charge is present on an initial configuration?
To address this question, we consider a family of initial configurations defined by

\begin{equation}
    \theta_n^\mathrm{init}\left(\theta_0\right)
    =\theta_0+\theta_n^Q,
\end{equation}
with $\theta_0\in\left[0,\pi\right]$. Given such an initial configuration, we investigate how the expectation value of the topological charge $Q$ evolves with Monte Carlo time $\tau_\mathrm{MC}$. Once the system has reached equilibrium, the expectation value of the topological charge is zero. We perform several independent runs to determine how the value $\left<Q\right>\left(\tau_\mathrm{MC}\right)$ depends on the value of $\theta_0$. The expectation is that, at least for $\theta_0=0$, the ISDB algorithm should reach $\left<Q\right>\left(\tau_\mathrm{MC}\right)\simeq 0$ for smaller $\tau_\mathrm{MC}$ than the standard algorithm. This is indeed what is observed in Fig. \ref{fig:relaxation}. 
In red we plot the evolution of the topological charge in the standard heat-bath algorithm, which takes the initial topological charge given to the system, and monotonically evolves it to zero. By construction of the algorithm, it does not depend on $\theta_0$. This is slower than the ISDB algorithm for an initial $\theta_0=0$, which is expected. The ISDB algorithm performs updates with an instanton background corresponding precisely to $\theta_0=0$ endowing it with an ability to quickly move on from that configuration. The more we deviate from that configuration, by increasing $\theta_0$, the more the algorithm gets slower. The performance becomes worse than the standard algorithm, for large times, once $\theta_0\gtrapprox0.2$, and keeps getting worse until $\theta_0=\pi/2$. By increasing more $\theta_0$, the configuration continues to differ more and more from the instanton background in the replica $\varepsilon=1$, but starts approaching the background instanton of the $\varepsilon=-1$, and the performance improves with it.

\section{Conclusion}

Topological freezing is a fundamental problem in lattice QCD, which extends to other physical theories. This typically occurs on local updating algorithms, which have difficulty climbing topological barriers. Solutions to this problem for certain models are known. For example, cluster algorithms can solve this problem for $O\left(N\right)$ models in $N-1$ dimensions \cite{wolff1989,wolff1990}. Proposals to solve this problem, when good cluster algorithms are not known, include change boundary conditions, master-field simulations, machine learning techniques, proposal of independent non-local updates or parallel tempering (some examples are \cite{luscher2011,luscher2018,kanwar2020,albandea2021,bonanno2022}).

Here we propose an alternative direction by exploring algorithms that break detailed balance. While this is being formulated as an alternative to the aforementioned methods, they are perfectly compatible. 
%The detailed balance condition is a powerful but restrictive condition to impose to the algorithm and, in principle, any algorithm can potentially be improved by breaking detailed balance in some way. In fact, we expect to be possible to construct more efficient algorithms capable of faster mixing and decorrelation of configurations in this way. 
We have explored a particular instance of breaking detailed balance, skewed detailed balance, properly modified to improve sampling of the topological charge in the 1-d $O\left(2\right)$ model, which we named Instanton-Skewed Detailed Balance algorithm. This algorithm operates by keeping instantons "in the background". By analyzing how the system relaxes from a configuration with topological charge, we observe that if the initial configuration is close enough to one of the possible instantons on the background, the system relaxes faster.
%Otherwise, its performance is worse than a standard algorithm. We emphasize that it is a question of performance and not of correctness: despite biasing the algorithm towards certain configurations, we still simulate the correct probability distribution.
The observed behavior suggests that the proposed algorithm, in its current form, is globally less efficient than the standard heat bath. However, for certain configurations that are similar to the background instantons, the converse is true. This gives hope that, by increasing the number of possible background configurations, one might be able to outperform the original algorithm altogether.

We emphasize that this is just a very specific way of breaking detailed balance. 
%It does so by utilizing a standard algorithm and increasing the chance of rejecting certain updates, depending on the value of a extra parameter (lifting variable). It would be interesting to construct an algorithm which does not rely on rejection of moves of a detailed balance algorithm, but then it becomes much harder to prove that we converge to the desired probability distribution. 
The results presented here suggest that skewed detailed balance can accelerate the sampling of different topological sections, but more research on it and other extensions are necessary to clarify a potential role in properly tackling the topological freezing problem.

\bibliographystyle{JHEP}
\bibliography{bible}

%\begin{thebibliography}{99}
%\bibitem{...}
%....
%\end{thebibliography}

\end{document}